\documentclass[aps,pra,reprint,onecolumn,notitlepage,eqsecnum,floatfix,nofootinbib]{revtex4-2}
\usepackage{amsmath}
\usepackage{color}
\usepackage{graphicx}
\usepackage{hyperref}
\usepackage{multirow}

\newcommand{\bq}{\begin{eqnarray}}
\newcommand{\eq}{\end{eqnarray}}
\newcommand{\bqn}{\begin{eqnarray*}}
\newcommand{\eqn}{\end{eqnarray*}}
\newcommand{\bqs}{\begin{subequations}}
\newcommand{\eqs}{\end{subequations}}
\newcommand{\bw}{\begin{widetext}}
\newcommand{\ew}{\end{widetext}}

\newcommand{\ZZ}{{\sf Z\!\!Z}}

\begin{document}
%%%%%%%%%%%%%%%%%%%%%%%%%%%%%%%%%%%%%%%%%%%%%%%%%%%%%%%%%%%%%%%%%%%%%%%%%%%%%%
%%%%%%%%%%%%%%%%%%%%%%%%%%%%%%%%%%%%%%%%%%%%%%%%%%%%%%%%%%%%%%%%%%%%%%%%%%%%%%
%%%%%%%%%%%%%%%%%%%%%%%%%%%%%%%%%%%%%%%%%%%%%%%%%%%%%%%%%%%%%%%%%%%%%%%%%%%%%%
\title{A Classical Luttinger liquid}

\author{Riccardo Fantoni}
\email{riccardo.fantoni@scuola.istruzione.it}
\affiliation{Universit\`a di Trieste, Dipartimento di Fisica, strada
  Costiera 11, 34151 Grignano (Trieste), Italy}

\date{\today}

\begin{abstract}
We propose a binary nonadditive Asakura-Oosawa-like mixture as an example for the classical 
limit of a Luttinger liquid. We determine the equation of state and structure of this mixture 
and discuss the paradoxical situation that one faces when working with a quantum liquid 
without a ground state. We then propose a new class of one dimensional classical fluids.
\end{abstract}

%\pacs{...}

\maketitle
%\tableofcontents
%%%%%%%%%%%%%%%%%%%%%%%%%%%%%%%%%%%%%%%%%%%%%%%%%%%%%%%%%%%%%%%%%%%%%%%%%%%%%%
\section{Introduction}
%%%%%%%%%%%%%%%%%%%%%%%%%%%%%%%%%%%%%%%%%%%%%%%%%%%%%%%%%%%%%%%%%%%%%%%%%%%%%%
\label{sec:intro}
In 1950 Tomonaga [Progr. Theoret. Phys. (Kyoto) {\bf 5}, 544
(1950)] published a theory of interacting fermions which was
soluble only in one dimension with the provision
that certain truncations and approximations were
introduced into his operators. Nevertheless he had
success in showing approximate boson-like behavior
of certain collective excitations, which he identified
as``phonons''. Today we would denote these as
``plasmons'', following the work of Bohm and Pines
[Phys. Rev. {\bf 92}, 609 (1953)]. 
Later, Luttinger [J. Math. Phys. {\bf 4}, 1154 (1963)] 
has revived interest in the subject
by publishing a variant model of spinless and massless
one dimensional interacting fermions, which
demonstrated a singularity at the Fermi surface.
Then Mattis and Lieb \cite{Mattis1965} corrected Luttinger results 
showing that there is a difference between
very large determinants and infinitely large
ones; they show that one of the important differences
is the failure of certain commutators to vanish
in the field-theoretic limit when common sense and
experience based on finite $N$ (here $N$ refers to the number of particles
in the field) tells us they should vanish. 

In this work we underline the Luttinger's paradox, of a many body system without a ground 
state, discussing the classical limit of his quantum liquid and propose a binary nonadditive 
Asakura-Oosawa-like mixture as a possible classical limit of a Luttinger liquid. We will 
then determine its equation of state and radial distribution function, exactly analytically. 
We therefore propose a new interesting class of one dimensional classical many body systems
as those that can be obtained as the classical limit of a Luttinger liquid.

%%%%%%%%%%%%%%%%%%%%%%%%%%%%%%%%%%%%%%%%%%%%%%%%%%%%%%%%%%%%%%%%%%%%%%%%%%%%%%
\section{The model}
%%%%%%%%%%%%%%%%%%%%%%%%%%%%%%%%%%%%%%%%%%%%%%%%%%%%%%%%%%%%%%%%%%%%%%%%%%%%%%
\label{sec:model}
Consider a Luttinger liquid \cite{Mattis1965}. Massless quantum indistinguishable 
particles of two species in one dimension. $N$ particles of species ``1'' with positions at 
$x_n\in[0,L)$ with $n=1,\ldots,N$ and $M$ particles of species ``2'' with positions at 
$y_m\in[0,L)$ with $m=1,\ldots,M$, with periodic boundary conditions on the segment $[0,L)$. 
The many particle system obeys to the following Hamiltonian
\bq
{\cal H}&=&{\cal H}_0+{\cal V}~,\\
{\cal H}_0&=&\sum_{n=1}^Np_{xn}-\sum_{m=1}^Mp_{ym}~,\\ \label{eq:interaction}
{\cal V}&=&2\lambda\sum_{n=1}^N\sum_{m=1}^MV(|x_n-y_m|)~.
\eq
where we adopt natural units $\hbar=c=1$ and 
$p_{xn}=-i\partial/\partial x_n$ and $p_{ym}=-i\partial/\partial y_m$ are the momenta of 
particles of species ``1'' and ``2'' respectively. In the original Luttinger liquid model 
\cite{Mattis1965} the particles are fermions and obey to the Fermi-Dirac statistics. But, for 
the time being and for the sake of simplicity, we will assume that they are distinguishable 
and obey to Boltzmann statistics. Then, if we define a function $W(x)$ as follows
\bq
dW(x)/dx={\rm sgn}(x)V(x)~,
\eq   
and we choose the following wave function
\bq \label{eq:wf}
\Psi=\prod_{n=1}^Ne^{ik_nx_n}\prod_{m=1}^Me^{iq_my_m}\exp
\left[i\lambda\sum_{n=1}^N\sum_{m=1}^MW(|x_n-y_m|)\right]~,
\eq
this is readily seen to obey the following Schr\"odinger's equation
\bq
{\cal H}\Psi=E\Psi~,
\eq
with just the unperturbed eigenvalues
\bq \label{eq:spectrum0}
E=\sum_{n=1}^Nk_n-\sum_{m=1}^Mq_m~.
\eq
The wave numbers are of the form
\bq \label{eq:wn}
k_n~~~{\rm or}~~~q_m=2\pi~{\rm integer}/L~,
\eq
as required by periodic boundary conditions. So the spectrum of ${\cal H}$ {\sl is the same as 
the one of ${\cal H}_0$ independent of the interaction $V(|x_n-y_m|)$}.

In the classical limit we should choose $p_{xn}\in[0,+\infty)$ and $p_{ym}\in(-\infty,0]$ so 
that the canonical partition function for the liquid at an absolute temperature 
$T=1/k_B\beta$, where $k_B$ is  constant, is well defined and given by
\bq \label{eq:pf1}
{\cal Z}(\{N,M\},L,T)&=&\int_0^Ldx^N\int_0^Ldy^M\int_0^\infty dp_x^N\int_{-\infty}^0 dp_y^M
\frac{e^{-\beta{\cal H}}}{(2\pi)^{N+M}N!M!}\\ \label{eq:pf2}
&=&\int_0^Ldx^N\int_0^Ldy^M\frac{e^{-\beta{\cal V}}}{(2\pi\beta)^{N+M}N!M!}\\ \label{eq:pf3}
&=&\exp(-\beta A)
\eq 
where $x^N=(x_1,\ldots,x_N)$, $y^M=(y_1,\ldots,y_M)$, $p_x^N=(p_{x1},\ldots,p_{xN})$, 
$p_y^M=(p_{y1},\ldots,p_{yM})$, and $A$ is the Helmholtz free energy.

Then for the ideal gas, ${\cal V}=0$, we find
\bq
\beta A&=&-\ln\left[\left(\frac{L}{2\pi\beta}\right)^{N+M}\frac{1}{N!M!}\right]~,\\
\beta P&=&-\frac{\partial\beta A}{\partial L}=\frac{N+M}{L}=\rho~,
\eq
with $\rho$ the density and $P$ the pressure of the ideal gas.

Imagine now a Hard Sphere (HS) potential with $\sigma_1=\sigma_2=0, \sigma_{12}=1$ as 
described in Sec. \ref{sec:ao}. Then, by following Eqs. (5.39)-(5.41) and (5.70) of Ref. 
\cite{Santosb}, one gets for the symmetric ($N=M$) binary mixture, the following 
compressibility factor
\bq \label{eq:eos}
Z=\frac{\beta P}{\rho}=1+\frac{\beta P}{1+e^{\beta P}}~.
\eq
The ideal gas equation of state (EOS) is obtained in the limits of low and high pressure.
The maximum value of $Z$ takes place at $\beta P\approx1.28$ and is $Z=\beta P\approx1.28$, 
i.e., at a density $\rho=1$. {\sl As expected, the EOS is different from that of the ideal 
gas}. 

So we reached a paradoxical situation in which the classical limit does not behave as the 
underlying more general quantum theory.

%%%%%%%%%%%%%%%%%%%%%%%%%%%%%%%%%%%%%%%%%%%%%%%%%%%%%%%%%%%%%%%%%%%%%%%%%%%%%%
\section{Solution of the paradox}
%%%%%%%%%%%%%%%%%%%%%%%%%%%%%%%%%%%%%%%%%%%%%%%%%%%%%%%%%%%%%%%%%%%%%%%%%%%%%%

In the path integral expression for the many body density matrix (see Eq. (2.12) of Ref. 
\cite{Ceperley1995}), Trotter's formula holds if the three operators ${\cal H}_0, {\cal V}$,
and ${\cal H}$ are self adjoint and make sense separately,
for example, if their spectrum is bounded from below. As we can immediately see from Eq. 
(\ref{eq:spectrum0}) this is not the case already for our ${\cal H}_0$ which does not 
admit a ground state.

First of all notice that if we restrict the wave numbers $k_n\in[0,+\infty)$ and 
$q_m\in(-\infty,0]$ then the eigenvalues of Eq. (\ref{eq:spectrum0}) admit 0 as lower bound.
But how can we achieve this restriction? Luttinger \cite{Mattis1965} proposed to assume that 
the particles are fermions 
\footnote{In this case the productorials in the wave function of Eq. (\ref{eq:wf}) must be 
replaced by determinants as follows $\det|e^{ik_nx_l}|$ and $\det|e^{iq_my_j}|$.}
and fill the ``infinite sea'' of negative energy levels (i.e., all states with
$k_n < 0$ and $q_m > 0$) with fictitious noninteracting particles. Only in this way one makes 
contact with a real massless fermion liquid with a ground state. Of course this requires 
working with an infinite number of particles. And the wave function $\Psi(x^N,y^M)$ becomes a 
functional $\Psi[\varphi_1(n),\varphi_2(m)]$ where $\varphi_\alpha(l)$ represents the 
position of the $l$th particle of species ``$\alpha$'', here $l\in\ZZ$ and $\alpha=1,2$. For 
example, we can adopt the convention of choosing for negative $l$ the fictitious fermions 
filling the negative energy levels and for positive $l$ the real fermions in the thermodynamic 
limit.

Of course in the classical limit the information on the indistinguishability of the particles 
is completely lost, the particles inevitably become labelable and distinguishable so that we 
have to artificially impose what may appear the additional artificial constraints 
$p_{xn}\in[0,+\infty)$ and $p_{ym}\in(-\infty,0]$. And this is what we did in the previous 
section.

From the quantum point of view, a general and inescapable concavity theorem states
that if $E_0(\lambda)$ is the ground-state energy in the presence of interactions 
(\ref{eq:interaction}) then
\bq
\partial^2 E_0(\lambda)/\partial\lambda^2<0~.
\eq
This inequality is incompatible with the previous result, viz. all $E =$ independent of 
$\lambda$, which was possible only in the strange case of a system without
a ground state. And this resolves the paradox.

It was first observed by
Julian Schwinger [Phys. Rev. Letters {\bf 3}, 296 (1959)] that the
very fact that one postulates the existence of a ground state
(i.e., the filled Fermi sea) forces certain commutators to be
nonvanishing even though in first quantization they automatically vanish. 
The ``paradoxical contradictions'' of which
Schwinger speaks seem to anticipate the difficulties in the
Luttinger model.

Luttinger mistake is well explained in Sec. III of Ref. \cite{Mattis1965} where it is clearly 
shown that the Fourier transform of the density operators $\rho_1(x)=\sum_n\delta(x-x_n)$ and 
$\rho_2(y)=\sum_m\delta(y-y_m)$, namely 
\bq
\rho_1(p)=\sum_{n=1}^Ne^{ipx_n}~,~~~\rho_2(q)=\sum_{m=1}^Me^{iqy_m}
\eq
which clearly commute in the finite number of particles liquid will no longer commute in the 
filled Fermi sea liquid in the thermodynamic limit, since
\bq \label{eq:do}
[\rho_1(-p),\rho_1(p')]=[\rho_2(p),\rho_2(-p')]=\delta(p-p')\sum_{-p<k<0} 1=
\delta(p-p')\frac{pL}{2\pi}~.
\eq
where for definiteness we chose $p\geq 0$ and $p'>0$ and we used Eq. (\ref{eq:wn}) to 
change the sum in the above equation (\ref{eq:do}), which takes into account the presence of 
the fictitious particles filling the Fermi sea, into an integral in the thermodynamic limit 
$L\to\infty$.
\footnote{For example, for species 1, the density operator is made of the sum of two parts, 
one for the real particles and one for the fictitious particles filling the Fermi sea. When we 
apply the commutator $[\rho_1(-p),\rho_1(p')]$ to the fluid state with $p\geq 0$ and $p'>0$ we 
find, from the cross terms, $(\sum_{k<0}-\sum_{k<-p})\sum_{n'}\exp[i(p'-p)x_{n'}]$ where the 
position of the fictitious particle at $x_n$ with momentum $-p<0$ can be taken equal to the
position of the real particle at $x_{n'}$ with momentum $p'>0$. The Dirac delta in Eq. 
(\ref{eq:do}) can then be understood thinking at the property 
$\sum_{n'}\exp[i(p'-p)x_{n'}]=\delta(p-p')$ in the thermodynamic limit 
$N\to\infty$.}

%%%%%%%%%%%%%%%%%%%%%%%%%%%%%%%%%%%%%%%%%%%%%%%%%%%%%%%%%%%%%%%%%%%%%%%%%%%%%%
\section{An Asakura-Oosawa-like model}
%%%%%%%%%%%%%%%%%%%%%%%%%%%%%%%%%%%%%%%%%%%%%%%%%%%%%%%%%%%%%%%%%%%%%%%%%%%%%%
\label{sec:ao}
As anticipated in Sec. \ref{sec:model} we can choose as a possible realization of a classical 
Luttinger liquid, a {\sl nonadditive} two component HS mixture where 
${\cal V}=\sum_{n=1}^N\sum_{m=1}^M\phi_{12}(|x_n-y_m|)$ with
\bq
\phi_{\alpha\gamma}(r)=\left\{\begin{array}{ll}
\infty     & r<\sigma_{\alpha\gamma}\\
0          & \mbox{else}
\end{array}\right..
\eq
Here, $\sigma_{\alpha\gamma}$ is the closest possible distance between the center of a sphere 
of species ``$\alpha$'' and the center of a sphere of species ``$\gamma$''. If we call 
$\sigma_\alpha=\sigma_{\alpha\alpha}$ the closest distance between two spheres of the same 
species ``$\alpha$'', it is legitimate to refer to $\sigma_\alpha$ as the
diameter of a sphere of species ``$\alpha$''. However, that does not necessarily mean that two
spheres of different type repel each other with a distance equal to the sum of their
radii. Depending on that, one can classify HS mixtures into additive or nonadditive:
\begin{itemize}
\item Additive HS (AHS) mixtures: $\sigma_{\alpha\gamma}=(\sigma_\alpha+\sigma_\gamma)/2$ for 
all pairs $\alpha\gamma$.
\item Nonadditive HS (NAHS) mixtures: 
$\sigma_{\alpha\gamma}\neq(\sigma_\alpha+\sigma_\gamma)/2$ for at least one pair 
$\alpha\gamma$.
\end{itemize}
Furthermore, the nonadditivity is said to be negative if 
$\sigma_{\alpha\gamma}<(\sigma_\alpha+\sigma_\gamma)/2$, while
it is positive if $\sigma_{\alpha\gamma}>(\sigma_\alpha+\sigma_\gamma)/2$.

As already announced in Sec. \ref{sec:model} we may realize a {\sl classical Luttinger liquid} 
by choosing a binary mixture where $\sigma_1=\sigma_2=0$ and $\sigma_{12}=1>0$. This is a 
NAHS with positive nonadditivity that describes an Asakura-Oosawa (AO) like model where not 
only polymers with each other but also colloids with each other are supposed not to interact 
\cite{Fantoni04a,Fantoni04b,Fantoni05a,Fantoni05b,Fantoni06a,Fantoni06b,Fantoni06c,
Fantoni07,Fantoni08a,Fantoni08b,Fantoni09a,Fantoni09b,Fantoni09c,Fantoni10a,Fantoni10b,
Fantoni11a,Fantoni11b,Fantoni11c,Fantoni11d,Fantoni11e,Fantoni12a,Fantoni12c,Fantoni13c,
Fantoni13d,Fantoni13e,Fantoni13f,Fantoni13h,Fantoni14a,Fantoni14b,Fantoni15a,Fantoni15c,
Fantoni16c,Fantoni16d,Fantoni17a,Fantoni17e,Fantoni23d}. So that, in the notation of Section 
\ref{sec:model}, we have $2\lambda V(x)=\phi_{12}(x)$
and in one dimension we should more properly talk about hard rods instead of HS.

As we can clearly see from Eqs. (\ref{eq:pf1})-(\ref{eq:pf3}) the non ideal 
configuration integral is not affected by the precise form of the 
kinetic energy term. This will affect the free energy but not the EOS \citep{Santosb}.
The nearest neighbor interaction condition requires that 
$\sigma_{\alpha\omega}\leq\sigma_{\alpha\gamma}+\sigma_{\gamma\omega}$ 
$\forall \alpha,\gamma,\omega$ which in a binary mixture reduces to 
$2\sigma_{12}\geq\max(\sigma_1,\sigma_2)$. So we will consider this 
nonadditive nearest neighbor hard rods binary mixture. As shown in Ref. \cite{Santosb} this 
fluid can be solved analytically exactly, both for the thermodynamics and for the structure, 
in the isothermal isobaric ensemble. 

Defining the molar fractions for the two species as $x_1=N/(N+M)$ and $x_2=M/(N+M)$ we find 
for the EOS
\bq \label{eq:feos}
Z=\frac{\beta P}{\rho}=1+\frac{\beta P[\sqrt{1+4x_1x_2(e^{2\beta P}-1)}-1]}{e^{2\beta P}-1}~.
\eq
This EOS reduces to the EOS given by Eq. (\ref{eq:eos}), in the symmetric case $x_1=x_2=1/2$.

\begin{figure}[htbp]
\begin{center}
\includegraphics[width=10cm]{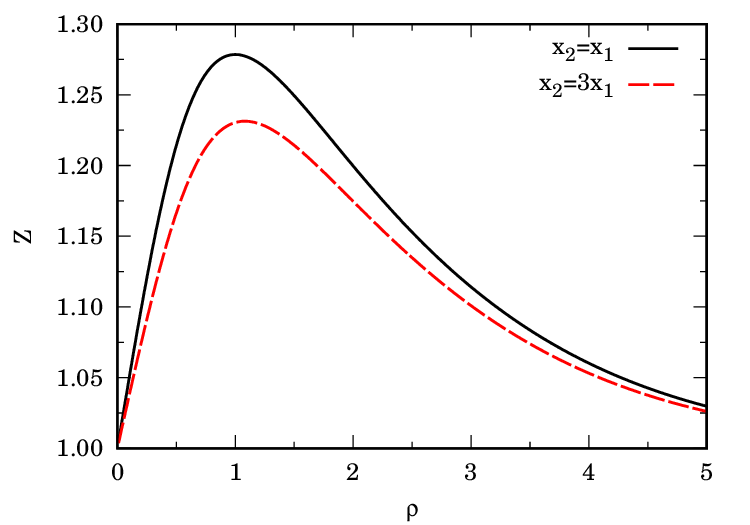}
\end{center}  
\caption{Compressibility factor $Z=\beta P/\rho$ for the nonadditive nearest neighbor hard 
rods binary mixture described in the text. $\beta=1/k_BT$ where $k_B$ is Boltzmann constant, 
$T$ is the absolute temperature, $P$ is the pressure, $\rho=(N+M)/L$ is the density, 
$x_1=N/(N+M)$ is the molar fraction of the particles of species ``1'', and $x_2=1-x_1$. 
The continuous line shows the result for the symmetric mixture and the dashed line the one for 
an asymmetric one.}
\label{fig:eos}
\end{figure}

The EOS of Eq. (\ref{eq:feos}) is shown in Fig. \ref{fig:eos} in a $(\rho,Z)$ plane for two 
choices of $x_1$.
The compressibility factor $Z$ has a non monotonic dependence since it starts from $Z=1$ 
at low density, reaches a maximum value ($Z=1.27846\ldots$ in the symmetric case at 
$\rho=1$) and then goes again to $Z=1$ in the high density limit. This is just because the 
state with the highest compressibility factor is made of $\min(N,M)$ pairs of ``1-2'' 
particles (corresponding to $\rho=1$ for the symmetric mixture) whereas in the high density 
limit the system segregates into $N$ particles of species ``1'' and $M$ particles of species 
``2'', separated by only a pair of ``1-2'' particles.

\begin{figure}[htbp]
\begin{center}
\includegraphics[width=10cm]{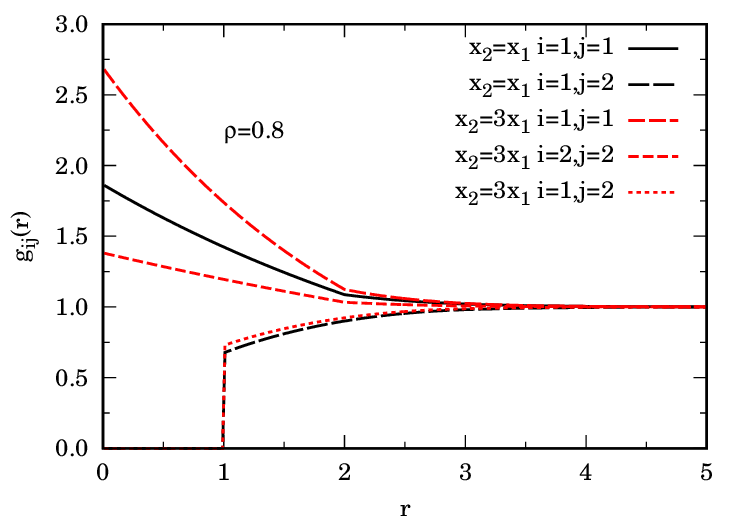}
\end{center}  
\caption{Partial radial distribution functions (like, $g_{11}$ and $g_{22}$, and unlike, 
$g_{12}$) as functions of $r=|x|$ for the nonadditive nearest neighbor hard rods binary 
mixture described in the text at a density $\rho=0.8$. Of course $g_{12}(r)$ vanishes for 
$r<\sigma_{12}=1$. We show the result for the two mixture already considered in Fig. 
\ref{fig:eos}. Clearly in the symmetric mixture $g_{11}=g_{22}$.}
\label{fig:gr}
\end{figure}

The structure of the mixture can be extracted from Eqs. (5.73)-(5.74) of Ref. \cite{Santosb}. 
In Fig. \ref{fig:gr} we show the partial radial distribution functions $g_{ij}(r)$ between a 
particle of species ``$i$'' and one of species ``$j$'', for the two mixture cases considered 
in Fig. \ref{fig:eos} at a density $\rho=0.8$, close to the maximum of $Z$. As we can see, 
there is a depletion of ``2'' particles and an abundance of ``1'' particles in the 
neighborhood of a ``1'' particle. We can also see a point of non derivability in the like 
functions at a distance $2\sigma_{12}=2$, which is due to the fact that one has a change in 
behavior when a ``2'' particle is allowed between two ``1'' particles. What is interesting of 
this AO-like model is that the asymptotic behavior of the total partial distribution functions 
$h_{ij}(r)=g_{ij}(r)-1$ is always monotonic with no oscillations!
\footnote{It can be shown that the dominant non zero pole of the Laplace transform 
of the partial radial distribution function (see Eqs. (5.37) in Ref. \citep{Santosb}) is the 
only real one. I thank Ana M. Montero for giving empirical evidence for this property.}

More generally we can say that the binary classical mixture limit of the Luttinger liquid 
must have non interacting like particles, i.e. $\phi_{\alpha\alpha}=0$ for $\alpha=1,2$. 
Moreover, the classical mixture is known to admit an exact analytic solution \cite{Santosb}
if in addition only {\sl nearest neighbor} interactions are possible, i.e.
\begin{itemize}
\item a particle is {\sl impenetrable} to a particle of different species, i.e. 
$\phi_{12}(r)=\infty$ for $r<{\cal I}$;
\item the pair interactions have {\sl finite} range, i.e. $\phi_{12}(r)=0$ for 
$r\geq{\cal R}$;
\item ${\cal R}\leq{\cal I}$.
\end{itemize} 
Therefore we conclude that only if the 1-2 interaction is of hard rod type 
(i.e., ${\cal R}={\cal I}=1$) in the classical Luttinger fluid can we expect that 
interactions are restricted to nearest neighbor.
\footnote{In the {\sl sticky limit}, since ${\cal R}\to 1$, it is not that evident 
that non nearest neighbor interactions are present, but intuitively one would say 
that this is still the case because if ${\cal R}-1$ is not strictly zero 
(even though we take the limit ${\cal R}\to 1$ at the end) you already have 
interactions beyond nearest neighbor. One would expect the sticky limit to be very 
close to a situation where ${\cal R}-1$ is very small but finite and the 
temperature is very low.}

%%%%%%%%%%%%%%%%%%%%%%%%%%%%%%%%%%%%%%%%%%%%%%%%%%%%%%%%%%%%%%%%%%%%%%%%%%%%%%
\section{Conclusions}
%%%%%%%%%%%%%%%%%%%%%%%%%%%%%%%%%%%%%%%%%%%%%%%%%%%%%%%%%%%%%%%%%%%%%%%%%%%%%%

We propose a realization of a classical Luttinger liquid and present its equation of state and 
structure. We discuss the Luttinger paradox and its solution which determines, through the 
classical limit, a new class of classical one dimensional fluids. We then discuss a case in 
which this class admits an exact analytic solution for the thermodynamics and the structure.

\acknowledgments
I thank A. Santos for many useful discussions during various stages of this work, and Ana M. 
Montero for providing empirical evidence for the property described in the footnote 3.
%%%%%%%%%%%%%%%%%%%%%%%%%%%%%%%%%%%%%%%%%%%%%%%%%%%%%%%%%%%%%%%%%%%%%%%%%%%%%%
\bibliography{cluttinger}
%\bibliographystyle{prsty}

%%%%%%%%%%%%%%%%%%%%%%%%%%%%%%%%%%%%%%%%%%%%%%%%%%%%%%%%%%%%%%%%%%%%%%%%%%%%%%
%%%%%%%%%%%%%%%%%%%%%%%%%%%%%%%%%%%%%%%%%%%%%%%%%%%%%%%%%%%%%%%%%%%%%%%%%%%%%%
%%%%%%%%%%%%%%%%%%%%%%%%%%%%%%%%%%%%%%%%%%%%%%%%%%%%%%%%%%%%%%%%%%%%%%%%%%%%%%
\end{document}